\documentclass[10pt]{article}

\usepackage{amsfonts}
\def\form#1{(\ref{#1})}
\def\Co{I \kern-.66em C}

\def\al{\alpha}
\def\b{\beta}

\def\L{{\mathcal L}}               
  \def\U{{\cal U}}  
\def\T{{\cal T}}                            
  
\def\H{{\mathcal H}}
\def\Q{{\mathcal Q}}              
\def\C{{\mathcal C}}

\def\V{{\mathcal V}}
\def\E{{\cal E}}

\def\U{{\mathcal U}}               

\def\F{{\mathcal F}}
\def\ram{\mathop{\longrightarrow}\limits}
\def\um{\mathop{=}\limits}

\def\DA{\stackrel{A}{D}}
\def\DG{\stackrel{\Gamma}{\nabla}}
\def\Dg{\stackrel{\gamma}{\nabla}}

\def\div{\mathop{\rm Div}\nolimits}

\def\Aut{\mathop{\rm Aut}\nolimits}

\def\Diff{\mathop{\rm Diff}\nolimits}

\renewcommand{\Re}{I\kern-.36em R}         

\newcommand{\be}{\begin{equation}}
\newcommand{\ee}{\end{equation}}
\newcommand{\ba}{\begin{eqnarray}}
\newcommand{\ea}{\end{eqnarray}}
\newcommand{\baa}{\be\left\{\begin{array}{l}}
\newcommand{\eaa}{\end{array}\right.\ee}

\def\QDE{\rule{2.5mm}{2.5mm}}
\def\CVD{$\phantom{'}$\hfill\QDE}

\newtheorem{Theorem}{Theorem}[section]

\newtheorem{Remark}[Theorem]{Remark} %
\newtheorem{Definition}[Theorem]{Definition} %
\newtheorem{Lemma}[Theorem]{Lemma} %
\newtheorem{Example}[Theorem]{Example} %
\newtheorem{Proposition}[Theorem]{Proposition} %
\newtheorem{Exercise}[Theorem]{Exercise}%

\title{Conserved Quantities from the Equations of Motion\\
\small (with applications to natural and gauge  natural
theories of gravitation)}
\author{M.\ Ferraris\thanks{E-mail:
ferraris@dm.unito.it},
M.\ Francaviglia\thanks{E-mail:
francaviglia@dm.unito.it}, M.\ Raiteri\thanks{E-mail:
raiteri@dm.unito.it}
\\
Dipartimento di Matematica, Universit\`a degli
Studi di Torino,\\
Via Carlo Alberto 10, 10123 Torino, Italy }

\date{}

\begin{document}
\maketitle

\begin{abstract}
We present  an alternative
field theoretical  approach  to the definition of conserved quantities,
 based directly on the field equations content of a Lagrangian theory  
(in the standard framework  of the Calculus of Variations in  jet
bundles). The contraction of the Euler--Lagrange equations   with 
Lie derivatives of the dynamical fields allows one to derive  a
\emph{variational Lagrangian} for any given set  of Lagrangian equations.  A
two steps  algorithmical procedure can be thence applied to the variational
Lagrangian in order to  produce a general expression for the variation of all
quantities which are (covariantly) conserved along the given dynamics. 
As a concrete example we test this new
 formalism on   Einstein's equations:
well known and widely accepted
  formulae for the
variation of the Hamiltonian and the variation of Energy for General
Relativity are recovered. We also consider the Einstein--Cartan
(Sciama--Kibble) theory in  tetrad formalism and as a  by--product  we
gain some new insight on  the Kosmann lift
in  gauge natural theories, which  arises when trying to
restore naturality in a gauge natural variational Lagrangian.

\end{abstract}

\section{Introduction}

A number of physically  reasonable 
 geometric definitions  of conserved
quantities in field theories  may be found in literature; in
the recent past
the issue  to define conserved quantities for
Lagrangian field theories  has been in fact   investigated  by many authors  
on the basis of different formalisms.  Just to mention a few of them we recall 
the  Lagrangian  method based on Noether's theorem
\cite{Lagrange,Wald,JuliaN,Noether,Trautman},  the Hamiltonian approach and
the symplectic methods
\cite{Anco,ADM,Boothultimo,BH,Waldsymp,Nester,HawHun,Kij,RT},
the  Hamilton--Jacobi--based  techniques
\cite{Booth,Mann94,BLY,BY},
and  the formulations which are  directly  based on 
field equations \cite{Torre,Silva, Rosen}. 
These are  just some  of the many references  which are most relevant  for our
purposes, whereby other important literature is quoted. Since a 
 satisfactory
review  is out of the scope of this paper  we apologize for not being able
to quote all authors  and sources  of information and we refer the reader to
the quoted papers and references therein.

In view of the fact that there exists in literature
 a widespread family  of somewhat unrelated definitions 
 it would be thence
rather important  to frame,  as much as possible,  all possible definitions 
under a unique and single method of construction. 
Very far  from  reaching such a fundamental goal, the  scope   of the present 
paper falls nevertheless into this line of thought. Our hope is  to
 shed at least some new light  on 
 a fully covariant  description of 
conserved quantities  in gauge natural field theories. We choose the 
so--called \emph{gauge natural}  framework \cite{Eck,Kolar} since it represents
a rich  geometrical  structure  which encompasses  all Lagrangian field
theories   relevant to fundamental physics (see also
\cite{Lorenzo2,Lorenzo,Godina,Matteucci}); this modern formalism is in fact
well suited to describe in a single mathematical context both natural
theories (the most famous of which is General Relativity) as well as gauge
theories together with  their possible  couplings with Bosonic and Fermionic
matter. This   setting becomes therefore a powerful tool  to frame  conserved
quantities  in a  geometric setting which is  general enough  to include all
relevant physical fields:  gauge natural theories play in fact a privileged
role   to this purpose since
 they  are
defined, from the very beginning, on the basis of the covariance properties
of the fields under the appropriate transformation groups. 
 
We point out that the gauge natural  formalism is
essentially a Lagrangian approach. Nevertheless field equations are 
 more fundamental than the Lagrangian itself
in the
description of dynamics.
Field equations, in fact,  rule out  those field configurations  which are
physically  admissible  and dictate their dynamical evolution; accordingly we
would prefer to  focus our attention 
 directly on the equations of
motion. 

However, it has to be remarked that the solutions of field equations  represent
just a small part  of the information that can be extracted  out of a field
theory. Other relevant information  is encoded  in the field symmetries;
  the knowledge on how fields are dragged along infinitesimal
generators of symmetries is  hence a further fundamental detail.  

Luckily enough,
the two main ingredients which have to enter into a satisfactory definition of
conserved quantities, namely the field equations content and the symmetry
information, can be joined together into a unique  structure. Field
equations are indeed described via the Euler--Lagrange morphism $e(L)$
which turns out to be a differential form on (some suitable jet prolongation 
of) the configuration bundle of the theory. The Euler--Lagrange morphism can
 be contracted with the Lie derivatives $\pounds_\xi y$ of the fields
$y$ with respect to infinitesimal generators $\xi$  of symmetries (which
define vertical vector fields). The resulting object $L':=-<e(L)\vert
\pounds_\xi y>$ turns out to be a horizontal form which  can be 
interpreted  as a new Lagrangian. From now on 
 we shall refer to $L'$ as \emph{the variational Lagrangian} (associated to
$L$ and $\xi$). The variational Lagrangian, for the information it encodes,  
is then in a good position to represent the fundamental object  
out of which we derive
 conserved
quantities: being $L'$ a Lagrangian  it
can in fact be handled by means of  the powerful tools of the Calculus of
Variations in  jet bundles. 

Our recipe to define
\emph{the variation} of conserved quantities will be thence developed in two
steps which are both canonically and algorithmically well--defined at the jet
bundle level. The first one is nothing but the \emph{first variational formula}
applied to the variational Lagrangian. The variation of $L'$, through a
well--known integration by parts procedure,  splits  canonically into  the
Euler--Lagrange  part (which vanishes because of   symmetry properties) plus  a
pure divergence term $\div
\F$. The so--called Poincar\'e--Cartan morphism  $\F$  which (non uniquely but
in a sense canonically) enters this divergence can be expanded as a linear
combination of  the coefficients of the vector field
$\xi$ (generating the flow of symmetries) together with their covariant
derivatives up to an appropriate (finite) order. The second step consists  in
implementing the so--called Spencer cohomology \cite{Robutti,Spencer}
through repeated  integrations by parts with respect to these covariant
derivatives. In this way we end up with a 
 ($m-2$) form $\U$,  called \emph{the potential},
where $m$ is the dimension
of spacetime,   which we choose to be a canonical representative, at the bundle
level, suited to define the variations of conserved quantities. These latter
quantities are indeed obtained by integrating the  pull--back
of
$\U$ along solutions on the appropriate 
($m-2$)--dimensional  domains of spacetime. 
 We point out that the possibility of  selecting a canonical representative for
the potential $\U$ is clearly an essential task, since different
representatives  may lead to different notions of conserved quantities.

Notice 
 that only the
\emph{variation} $\delta\Q$ of a conserved quantity, rather than the conserved
quantity $\Q$ itself, is here defined.  However, one can  argue that 
$\delta\Q$ rather than $\Q$ is the truly fundamental object. Physically
speaking, in fact, conserved quantities are not absolute, since
only differences of physical observables are endowed with a direct physical
meaning. Therefore  one has  to somehow fix   a
``reference point'' (e.g. a background) to calculate them.
However, in the absence of a linear  structure on the space of  fields  there
is no canonical choice for such a  ``reference point''. Moreover, the
formal on--shell integration of
$\delta
\Q$  depends on how we move along a curve in  the space $\E$ of
solutions. Starting from a given solution $\varphi_0$, different deformations
$\delta $  correspond to different paths in $\E$ passing through the same point
$\varphi_0$,  so that $\delta
\Q$ measures the variation of $\Q$  when we move from $\varphi_0$ to a  nearby
solution in a given direction. Roughly speaking  the choice   of such a
direction corresponds to a control mode of the fields at the boundary which
 is based on physical grounds (e.g. micro--canonical or
grand--canonical ensembles) and it is  related to the asymptotic behaviour of
specific solutions and to the specific physical observable one is aimed to
determine (see, e.g., \cite{Anco,Barnich,BY,Nester,forth,Silva,Kij}).  We then
believe that, inside  a formalism which we want to be as general as possible,
both physical and mathematical reasons suggest  to us that  only variations
$\delta
\Q$ should be reasonably  considered.\vspace{.8truecm}

Finally, we point out another property that  a viable definition of  conserved
quantities  has to fulfill (and according to which we have developed 
the present formalism). Conserved quantities  are mathematically defined
objects which are built out of fields with the purpose of  extracting
information  about  the physical properties of all solutions; they represent
indeed physical observables such as mass, energy, momentum, angular momentum
or gauge charges. For this reason  the mathematical definition of conserved
quantities, whatever formalism we implement to obtain it,    
   must eventually get rid of the
mathematical structure we started from and, in the meanwhile,  it must not
depend on the specific configuration variables we have selected  to describe
the solution itself. Let us better  explain this ``philosophical'' viewpoint by
considering as an example the theory of gravitation. It is well-known that
there exist different Lagrangian formulations of gravity (e.g. purely
metric \cite{Hilbert15}, metric affine \cite{Einstein25}, purely affine
\cite{Einstein23}, tetrad formulations
\cite{Kibble}, Ashtekar variables \cite{A}, Chern--Simons formulation
\cite{Ach} and so on); these are described  by profoundly different
Lagrangians which in many cases  are ``equivalent'' in the sense that  they
determine  equivalent field equations even if they do not merely differ   for
the addition of divergence terms. From a geometric viewpoint  these are really
different theories  since they are based on  different configuration spaces
and  involve different dynamical fields. Nevertheless, all the aforementioned
theories, under  suitable regularity conditions, are equivalent on shell
(i.e. they generate  essentially identical or at least  isomorphic spaces of
solutions, possibly under appropriate gauge reductions).
Basically, there exists  a rule
(which can be many--to--one) which maps  a solution of one theory  into a
solution of another.\footnote{For example purely metric, metric affine and 
purely affine theories are all related by a generalized Legendre transformation
and the quoted map among solutions is just given by the Legendre map
\cite{Kjj}} For each different formulation of gravity we expect then a
different setting for the definition of conserved quantities. Nevertheless it
is physically desirable that all different  definitions  of conserved
quantities turn eventually out to coincide on--shell; otherwise we  may risk
to obtain   a paradoxical conceptual result. Indeed, if 
on--shell equivalent gravitational theories admit 
 solutions that
correspond   to the same spacetime, all physically meaningful observables must
be related to the spacetime itself and should  not depend on which particular
variable we have used to describe it. For example, the energy enclosed in a
(bounded) region surrounding a black hole solution must depend solely on the
 solution and not on the specific configuration  variable (e.g.
metric, tetrad or  connection) we initially choose to describe it.

The formalism developed in this paper partially avoids such a possible 
paradox. In fact, it is based on the variational Lagrangian
which  is obtained via contraction of  the equations of motion
with the Lie derivatives of fields. If  any rule exists  to unambiguously
map the equations of motion and the Lie derivatives  of one theory  into the
corresponding objects  of another theory, then  there exists, of course,  an
unambiguous  correspondence between the variational Lagrangians, too. 
Accordingly,  also the conserved quantities
of the first theory will be  mapped into the conserved quantities of
the second,  so that, on--shell, they will eventually give rise  in both
approaches to the same numerical value for each conserved integral. 

However, we have to stress that 
there exist Lagrangian theories which admit different symmetry groups even if
they are equivalent as far as their field equations content is concerned. For
instance,  purely metric, metric affine and  purely affine theories  are
natural formulations of gravity and they all admit the group  of spacetime
diffeomorphisms as  the ``natural'' symmetry group. On the other hand  tetrad
or Chern--Simons formulations of General Relativity are  truly gauge 
natural formulations so that  they admit  a much larger group of symmetries,
which,  besides spacetime
diffeomorphisms, encompasses also gauge transformations.

In general, in gauge natural theories it happens that pure spacetime
transformations cannot be globally isolated (in non trivial topologies) from
gauge transformations. This fact reverberates then into the indeterminacy  in
defining the Lie derivative of gauge natural objects with respect to arbitrary 
spacetime vector fields (while such an indeterminacy does not occur for
natural objects).\footnote{E.g., spinors, which are truly gauge natural
objects cannot be classically Lie--dragged  along generic spacetime  vectors,
but a classical notion  of Lie derivative  exists along  Killing vectors,
i.e. infinitesimal symmetries  of the spacetime metric. A general definition 
of Lie derivative of spinors can be achieved  \emph{only} in the gauge
natural formalism (see
\cite{Lorenzo,Godina,Matteucci}). }
 Many lifts of the same
vector field can be thence  chosen: they originate  different Lie
derivatives and, accordingly,  different variational Lagrangians and different
conserved quantities. This fact raises a very interesting issue about
symmetries (which we will face up   in the last section).  Namely, albeit we 
have a correspondence  between equations of motion  we,
\emph{a priori}, lack  a strict correspondence between symmetries.  
Nevertheless we claim  that, \emph{a posteriori}, also the latter
correspondence  can be gained in our approach and such a goal  is
simply obtained
 by imposing the equivalence of the variational Lagrangians. In this
way  a preferred lift  of spacetime vector fields can be ruled out in gauge
natural theories: it is the lift which restores the naturality in the  gauge
natural variational Lagrangian.
\vspace{.8truecm} 

The present paper is organized as follows. In section \ref{Natural and Gauge
Natural Theories}, in order   to make  the paper self--contained,  we
shortly  review the geometric framework of natural and gauge natural theories.
In  section
\ref{section3} we  present the theoretical formulation of the ``two steps
procedure'' which  leads to the definition of the variation of conserved
quantities. The result  achieved here is tested  for the purely
metric formulation of General Relativity (in section \ref{section4}) and 
  for the gauge natural tetrad formulation of General Relativity (in
section \ref{section5}). No discrepancies are
observed between  the two different formulations owing  to the specific form of
the variational Lagrangians. The last section (section \ref{KosmannS}) is
finally devoted to the analysis of Einstein--Cartan theory in the gauge natural
framework. This theory is  a generalization of Einstein's theory and, in
vacuum, it becomes on--shell equivalent to it.  We  have chosen to investigate
this theory  since it clearly features the indeterminacy we mentioned above 
when defining  conserved quantities in gauge natural theories with respect to
spacetime vector fields. It will become however clear how this indeterminacy
can be a posteriori eliminated by restoring naturality in the  gauge natural
variational Lagrangian. This procedure selects in the specific example the
so--called \emph{generalized  Kosmann lift}  as the  preferred lift to deal
with. As a  by--product  we  gain some new insight on  the mathematical
justification of the Kosmann lift previously introduced  in
\cite{Lorenzo,Godina,Matteucci} in the domain of gauge natural theories as a
generalization  of an {\it ad hoc} procedure  introduced by Kosmann  in
\cite{Kosmann} to Lie drag spinors.


\section{Natural and gauge natural theories}
\label{Natural and Gauge Natural Theories}
Natural theories  geometrically  formalize  the physical
principle of general covariance. According to \cite{Lagrange,Kolar} we say that
a field theory is
\emph{natural} when: 
\begin{enumerate}
\item[A:] the
configuration bundle is natural;
\item[B:] the Lagrangian describing the theory is 
 natural.
\end{enumerate}
Item A means  that spacetime diffeomorphisms  can be functorially lifted
to the configuration bundle $Y$ (i.e. the  space where the  dynamical
fields $\varphi$  take their values). 
Roughly speaking, we know how fields transform under changes of coordinates in
 spacetime. As a consequence of naturality, for each spacetime vector field
$\xi$ there exists a canonical lift  $\hat
\xi$ on $Y$ (which projects onto $\xi$). Using the natural
lift
$\hat
\xi$  it is   then  meaningful to consider Lie derivatives of fields with
respect to spacetime vector fields by setting $\pounds_\xi
\varphi:=\pounds_{\hat\xi}
\varphi=T\varphi \circ\xi -\hat\xi\circ \varphi$.

Item B means that all spacetime transformations, once  lifted on  $Y$, 
are symmetries. Accordingly, each lift $\hat \xi$ is an infinitesimal
generator of symmetries.

Obviously all theories based on  diffeomorphism invariant  Lagrangians
which depend  on tensor fields, tensor densities and/or linear connections
are natural theories according to the given definition. 

In order to encompass into a unique geometric formalism  theories
admitting both diffeomorphism invariance as well as gauge symmetries one is 
led to introduce \emph{gauge natural theories}. In gauge natural theories we
assume that there exists a principal bundle $(P,M,p;G)$, called \emph{the
structure bundle}, where  all information concerning symmetries is encoded;
 gauge natural theories admit  the group
$\Aut(P)$ of all automorphisms  of the structure bundle as group of symmetries
(pure gauge symmetries correspond to vertical automorphisms). This amounts to
say that a theory is gauge natural (see \cite{Lorenzo,Lorenzolibro,Kolar} for 
a deeper geometric insight) when: 
\begin{enumerate}
\item[A:] the
configuration bundle is gauge natural;
\item[B:] the Lagrangian describing the theory is  gauge natural;
\item[C:] a linear connection $\Gamma$ and a principal connection $A$ on the
structure bundle  can be built out of the dynamical fields.
\end{enumerate}
\label{G_N}
Item A is a pure geometric requirement: it means  that the automorphisms of
the structure bundle  $P$ functorially induce automorphisms on the 
configuration bundle $Y$.  Accordingly, 
 projectable vector fields $\Xi_P$ on $P$ canonically induce vector fields 
$\Xi$ on $Y$. Item B is instead of dynamical nature since it implies that all
such induced vector fields
$\Xi$ are infinitesimal generators of symmetries. Therefore we should be
somehow able  to  associate conservation laws to each of them  leading to
quantities which are eventually  physically interpretable as observables.
  In doing that we shall need  the    Lie derivatives of fields with
respect to  vector fields $\Xi_P$, which  are
defined   by setting
\be
\pounds_{\Xi_P}
\varphi:=\pounds_{\Xi}
\varphi=T\varphi \circ\xi -\Xi\circ \varphi
\ee where $\xi$ denotes the
projection of $\Xi_P$ onto  spacetime $M$. Notice however that the group
$\Diff(M)$ is not canonically embedded into
$\Aut(P)$. We know how fields transform as a consequence of a
transformation  in $P$  but  we  do not know how fields transform
under change of coordinates in  spacetime. In other words Lie derivatives
with respect to spacetime vector field cannot be, at least a priori,  defined
for a generic $\xi$. 

Finally, item C is a technical requirement: the two \emph{dynamical}
connections
$\Gamma$ and $A$ are the mathematical tools  which are necessary  to  provide
covariance at each step of the geometric formalism used to generate 
conserved quantities.

\begin{Remark}{\rm We stress that despite in gauge natural theories there
exists no natural way to  define the action of diffeomorphisms  on the
dynamical fields there exist, however, \emph{many} global (but not canonical)
ways to lift  spacetime vector fields up to  the structure bundle. One of this,
for example, is \emph{the horizontal lift} defined through the dynamical
connection
$A$. Even if this lift apparently seems to be the most ``natural'' way to
define the lift of vector fields, it has  been  nevertheless shown
\cite{Ultimo} that it does not lead to physically acceptable  values for
conserved quantities in physical applications  so that eventually one has to
resort to some other lift. In the applications we shall deal  with, it will be
shown that the generalized Kosmann lift is the most ``natural'' one; see
section
\ref{KosmannS}. \CVD}
\end{Remark}

\begin{Remark}{\rm We remark that, as far  as the geometric formalism
for conserved quantities is   concerned, we are not so much 
interested in the Lagrangian  but rather in  the variational equations
ensuing from it. Therefore we could have weakened  item B by just requiring
that the dynamical equations describing the physical system are (gauge)
covariant. Nevertheless, starting from a set of covariant equations of
variational nature  it is always possible, at least in principle, to build out
a covariant family of  Lagrangians depending on a (dynamical) background; see
\cite{Tapia} (see also
\cite{Ultimo,Borowiec} where covariant Lagrangians for Chern--Simons theories
are exhibited).  In the sequel we shall implicitly assume that any one of
these  Lagrangians has been already selected, even if we shall not be
interested into its explicit form.\CVD}
\end{Remark}

\section{Conserved Quantities from the Equations of Motions}
\label{section3}
Let us consider a gauge natural theory geometrically
described through a gauge natural bundle
$(Y,M,\pi)$ and dynamically defined by   a
$k$--order  Lagrangian $L: J^k Y\ram \Lambda^m(M)$. In terms
of fibered coordinates $(x^\mu, y^i)$ on $Y$ we locally have 
$L=\L( j^k y)
\, ds$ where $\L$ is the Lagrangian density,
$ds=dx^1\wedge\dots\wedge dx^m$ is the standard (local) volume form
on $M$ and $(j^k y)$ stands for $(y^i, y^i_\mu,\dots, y^i_{\mu_1\dots\mu_k})$,
i.e. the set of partial derivatives  of fields up to order $k$ included. We
shall denote by
$J^k Y$ the
$k$ jet prolongation of
$Y$, by  
$V(J^kY)$ its  vertical tangent bundle and by  
$V^*(J^kY)$
 the vector bundle  dual to $V(J^kY)$.
Notation  and definitions  in this section  follow closely
\cite{Lorenzolibro},  to which we refer the reader  for a full
treatment  and further details.

\emph{A variation}
is  a vertical vector field $X$ on the configuration
bundle, which  can be locally described as $X=X^i{\partial
\over
\partial y^i}=\delta y^i{\partial
\over
\partial y^i}$;  physically speaking  it describes a
one--parameter deformation of the dynamical fields. Accordingly, we can
consider the variation
$\delta_X L$ of  the Lagrangian  along  the flow of (the prolongation
of) the vector field $X$. It is well--known (see
\cite{Lagrange,antichi,Trautman}) that  each Lagrangian $L$ induces  a unique
(global) morphism, called {\it the Euler-Lagrange morphism}
\be
e(L):J^{2k}Y\ram V^*(Y)\otimes \Lambda^m(M)\label{2}
\ee
together with a family of (global) morphisms (which depend on the Lagrangian 
and possibly on a
connection $\Gamma$ on $M$) called {\it Poincar\'e-Cartan morphisms}
\be
\F(L,\Gamma):J^{2k-1}Y\ram V^*(J^{k-1}Y)\otimes \Lambda^{m-1}(M)
\ee
The Euler-Lagrange morphism and the Poincar\'e-Cartan morphisms are in fact
defined so that the so-called {\it first-variation formula} holds for any
deformation $X$ on $Y$:
\be
\delta_X L=\><e(L)\>\vert\> X>+\>\div<\F(L,\Gamma)\>\vert\>
j^{k-1}X>\label{FVF}
\ee
where $<\,\vert\, >$ denotes the canonical pairing
between differential forms and vector fields (in our case between elements of
$V^*(J^hY)$ and elements of $V(J^h Y)$). The {\it
formal divergence operator on forms} is defined by
\be
\div(f)\circ j^{k+1}\varphi= d(f\circ j^k\varphi),
\qquad\qquad f:J^kY\ram \Lambda(M)	
\ee
$d(\cdot)$ being the exterior differential operator on forms and
$\Lambda(M)\equiv\oplus_k \Lambda^k(M)$ denoting  the bundle of forms over
$M$.
The  Poincar\'e-Cartan morphisms are uniquely defined only for $k=0$ or $k=1$;
for
$k=2$ 
they are not unique but still there is a canonical choice which is independent
on any connection $\Gamma$. For $k>2$ the Poincar\'e-Cartan morphisms
strongly depend on the choice of a linear connection; see \cite{Ferraris}.
Nevertheless, in gauge natural theories the existence of a dynamical linear
connection
$\Gamma$ is axiomatically required (see item C of the definition); in this
case  we shall implicitly assume that the Poincar\'e-Cartan morphism entering
 the first variation formula (for $k>2$) is  the one induced  by this
preferred  connection $\Gamma$. For this reason, from now on  we shall omit
to indicate in the notation the dependence on
$\Gamma$ of    
$\F(L,\Gamma)$ and we shall simply write $\F(L)$ .


The Euler--Lagrange morphism \form{2}, which can  can be locally written as 
\be
e(L)=e_i(j^{2k} y)\, dy^i\otimes ds\nonumber 
\ee
 encodes the information relative to  the equations
of motion for the dynamical fields. A \emph{critical
section} (or a solution) is a section $\varphi:M\ram Y$,
locally described as $\varphi: x^\mu\mapsto (x^\mu,
y^i=\varphi^i(x))$, the prolongation of which belongs to the
kernel of the  Euler--Lagrange morphism, i.e.:
\be
e(L)\circ j^{2k} \varphi=0\quad \Longrightarrow \quad
e_k(\varphi^i, d_\mu
\varphi^i,\dots, d_{\mu_1\dots \mu_{2k}}\varphi^i)=0
\ee
Now,  let us denote by
$\Xi_P$ a projectable vector field on the relevant principal
bundle
$(P,M,p)$ of the theory. It locally reads as $\Xi_P= \xi^\mu
\, \partial_\mu +\xi^A\,\rho_A$ (where $\rho_A$ ia
a local basis for right invariant  vertical vector fields 
on $P$).  It canonically induces a vector field $\Xi$
 on the configuration bundle  which is, by the
very definition  of gauge natural theory (see the previous section), an
infinitesimal generator of symmetries. Through the vector field $\Xi$ we
can define the (formal) Lie derivative $\pounds_{\Xi} y$ of the
fields. Since $\pounds_\Xi y:J^1Y \ram V(Y)$ takes value into the vertical
bundle of $Y$ it is meaningful to consider the contraction 
\be
L'(L, \Xi)= -<e(L)\,\vert\, \pounds_\Xi
y>:J^{2k} Y\ram \Lambda^m  (M)\label{Lprimo}
\ee
which  defines  a horizontal form on the bundle $J^{2k}Y$  which we shall
call, from now on, \emph{the variational Lagrangian}.

 Since we are assuming the
configuration bundle to be  a gauge natural bundle 
 the Lie derivative $\pounds_\Xi
y$ entering into the definition  of the variational Lagrangian can be
written as a linear combination of symmetrized covariant derivatives 
with respect  to the dynamical connections $(\Gamma, A)$; see
\cite{Remarks,Robutti}. Thereby expression  \form{Lprimo} can be locally
written as
\ba
L'(L, \Xi)&=&\left\{
W^{}_\mu\xi^\mu+W^{\rho_1}_\mu\nabla_{\rho_1}\xi^\mu+\dots
+W^{\rho_1\dots\rho_{r}}_\mu\nabla_{(\rho_1\dots\rho_{r})}\xi^\mu\right.
\label{RRR}\\
&&\left.+W^{}_A\xi^A+W^{\rho_1}_A\nabla_{\rho_1}\xi^A+\dots
+W^{\rho_1\dots\rho_{s}}_A\nabla_{(\rho_1\dots\rho_{s})}\xi^A
\right\}\, ds\nonumber
\ea
where 
$W^{}_\mu, W^{\rho_1}_\mu$, $\dots$,
$W^{\rho_1\dots\rho_{s}}_\mu$, $W^{}_A$, $W^{\rho_1}_A$, $\dots$,
$W^{\rho_1\dots\rho_{r}}_A$ are {\it tensor densities} with respect to
automorphisms of the structure bundle (the pair $(r,s)$ is called  \emph{the
order} of  the gauge natural bundle; see \cite{Kolar}); here and in the sequel
round brackets around indices denote symmetrization.

Whenever we have such a linear combination we can perform covariant integration by parts
to obtain for $L'(L, \Xi)$ an equivalent canonical  expansion under the form:
\be
L'(L, \Xi)={\cal B}(L,\Xi)+\div\>\tilde\E(L,\Xi)
\label{11}
\ee
where the quantity $\tilde\E(L,\Xi)$ is usually called the {\it reduced
current} and vanishes on--shell, while  the quantity 
${\cal B}(L,\Xi)$ is linear in $\Xi$ and it turns out to be
identically vanishing along any section (see
\cite{Lagrange,Robutti}). The identities 
\be
{\cal B}(L,\Xi)=0
\ee
are called {\it generalized Bianchi identities} (indeed they reduce to the
usual Bianchi identities in General Relativity and in gauge theories).
Hence equation \form{11} gives rise to a conservation law:
\be
\div \tilde\E(L,\Xi)=L'(L, \Xi)\simeq 0\label{trivial}
\ee
where $\simeq $ denotes equality on--shell. Nevertheless it is clear  that
\form{trivial} is a \emph{ trivial} conservation law owing to the fact that
the reduced current $\tilde \E$, which is  built out of the coefficients
\form{RRR} of the Euler--Lagrange morphism  together with their covariant
derivatives,  is, as we said, vanishing on--shell. Hence equation
\form{trivial}     is of no utility  by
itself to describe physical properties of any given solution. Nevertheless, 
the variational Lagrangian   becomes the starting point to 
define algorithmically the quantities we are interested in.
The formalism basically
develops in two steps.\vspace{.8cm}

{\noindent \bf First step: integration by parts with respect to $X$}.

 \noindent Let us
consider
 a vertical vector field $X$ on the configuration
bundle, locally  given as $X=X^i{\partial
\over
\partial y^i}=\delta y^i{\partial
\over
\partial y^i}$. By treating \form{Lprimo} as a new Lagrangian we
can consider the variation $\delta_X L'$ and, accordingly, we can make again
use of the first variational formula \form{FVF}. Notice that
$L'$ in general depends on the fields $y^i$ together with  their derivatives up
to some  order
$h\le 2k$, but it also
depends on the variables $\xi^\mu$ and $\xi^A$ together with their derivatives
up to the orders 
$r$ and $s$, respectively; see \form{RRR}.\footnote{The inequality $h\le 2k$
is due to the fact that a Lagrangian of order $k$ can give rise to field
equations of order lower than $2k$; see, e.g. General Relativity or Lovelock
metric theories
\cite{CF} where $k=2$ while  field equations are second order only.}
 In the variation $\delta_X L'$ the components
$\xi^\mu$ and
$\xi^A$ are  independent on fields and thereby are kept fixed, i.e. $\delta_X
\xi^\mu=\delta_X
\xi^A=0$. In the sequel we shall also   be interested  in the case in which 
the components $\xi^A$, via some suitable ``lift'', are built out of the
dynamical fields $y^i$ and  the coefficients
$\xi^\mu$ together with their derivatives up to some fixed order, i.e.
$\xi^A=\xi^A(j^a y, j^b \xi^\mu)$. When such a  dependence is inserted back
into
\form{Lprimo} we end up with 
a variational Lagrangian which depends only
on the dynamical fields  and the components $\xi^\mu$. Thereby 
the whole formalism we are going to develop will  simply work by setting $\xi^A
=0$.

 The first variation
formula applied to the Lagrangian
$L'$ reads now  as follows:
\ba
\delta_X L'&=&<e(L')\,\vert\, X> + \div \F (L',X)\nonumber\\
&=& \div \F (L',X)\label{deltaL'}
\ea
where $\F (L',X)$ stands for $<\F (L')\>\vert\> j^{p}
X>$ for some suitable $p$. 
The latter equality in \form{deltaL'} is due to the fact that,
being the Lagrangian $L'$ a total divergence (see \form{trivial}) its
associated Euler--Lagrange morphism  $<e(L')\,\vert\, X>=<e(\div \tilde
\E)\,\vert\, X>$ vanishes identically. Moreover the variation of 
\form{trivial} turns out to be a pure divergence, i.e.:
\be
\delta_X L'=\div \delta_X\tilde\E(L,\Xi)\label{xxx}
\ee
Hence, comparing \form{deltaL'} with \form{xxx} we obtain a \emph{strong}
conservation law (i.e. a conservation law which holds true off--shell):
\be
\div \left\{\F (L',X)-\delta_X\tilde\E(L,\Xi)\right\}=0\label{F-E}
\ee
In particular, if the variation $X$ is a solution of the linearized field
equations, i.e. it is tangent to the space of solutions, then
$\delta_X\tilde\E(L,\Xi)=0$  and we obtain a \emph{weak} conservation law:
\be
\div \F (L',X)\simeq 0\label{12}
\ee\vspace{.5cm}

{\noindent \bf Second step: integration by parts with respect to $\Xi$.}

\noindent We
shall now show that the $(m-1)$ form 
$\F (L',X)-\delta_X\tilde\E(L,\Xi)$ is not only closed but it is also exact
(off-shell) and we shall explicitly exhibit a global representative
for its potential. This is a result which does not depend on the topology of
spacetime and it does not depend on any given solution. Indeed all calculations
will be performed 
on some  jet prolongation of  
the configuration bundle $Y$.
Only at the end of the calculations we shall pull--back the results 
along
sections of the appropriate  jet bundle obtaining in this way
differential forms on the base manifold. Moreover the global character of the
results we shall obtain will follow directly from  the fact that the algorithm
 developed  fulfills  the covariance property at each step of its
construction. 

First of all let us consider the morphism $\F (L',X)$\label{pagexx}. Since
$L'$ depends on
$j^{h}y$, $j^r\xi^\mu$ and $j^s \xi^A$, with $h\le 2k$,
and the expression in the right hand side of
\form{deltaL'} is obtained, in general,  through $h$ integration by parts, the
quantity $\F (L',X)$ depends linearly on the independent components
$\xi^\mu$, $\xi^A$ together with  their derivatives up to the orders,
respectively,
$h+r-1$ and $h+s-1$. Therefore it can be  written as
\ba
\F (L',X)&=&\left\{
\F^{\rho}_\mu\xi^\mu+\F^{\rho\rho_1}_\mu\nabla_{\rho_1}\xi^\mu+\dots
+\F^{\rho\rho_1\dots\rho_{h
+r-1}}_\mu\nabla_{(\rho_1\dots\rho_{h
+r-1})}\xi^\mu\right.\label{fff}\\
&&\left.+\F^{\rho}_A\xi^A+\F^{\rho\rho_1}_A\nabla_{\rho_1}\xi^A+\dots
+\F^{\rho\rho_1\dots\rho_{{h +s-1}}}_A\nabla_{(\rho_1\dots\rho_{{h
+s-1}})}\xi^A
\right\}\, ds_\rho\nonumber
\ea
where $ds_\rho=\partial_\rho\rfloor ds$.
The coefficients in \form{fff} are {\it tensor densities} which depend on the
dynamical fields $y^i$ together with  their variations $X^i=\delta y^i$ up to
some (finite) order. Notice that
covariant derivatives can be  defined with respect to the dynamical
connections (see item C of the definition). 
Notice also that  the coefficients $\F^{\rho\rho_1\dots\rho_{p}}_\mu$ and 
$\F^{\rho\rho_1\dots\rho_{q}}_A$
are
of course symmetric  in the indices $\rho_i$ but not with respect to the whole
set  of upper indices. However,
whenever we have  a linear
combination of the kind
\form{fff} we can perform covariant  integration by parts to obtain for the
same quantity  an equivalent linear  expansion  the coefficients of which are
all symmetric  with respect to upper indices, while the integrated terms  are
all pushed into a formal divergence. In this way expression \form{fff} can be
recasted as follows:
\footnote{\label{spencer}
For the sake of completeness of this paper we give the explicit formula for the
 decomposition in the case $h+r-1=2, s=0$ and
$\Gamma^\alpha{}_{\mu\nu}$ symmetric. We have in this case:
\ba
&&\tilde\F^{\alpha}_\mu=\F^{\alpha}_\mu+f^{\alpha}_\mu-\nabla_\beta
(\F^{[\alpha\beta]}_\mu+
f^{[\alpha\beta]}_\mu)\nonumber\\
&&\tilde\F^{(\alpha\beta)}_\mu=\F^{(\alpha\beta)}_\mu+f^{(\alpha\beta)}_\mu
\label{FFTILDE1}\\
&&\tilde\F^{(\alpha\beta\gamma)}_\mu=\F^{(\alpha\beta\gamma)}_\mu\nonumber
\ea
with
$f^{\alpha}_\mu=1/3 \F^{[\sigma\beta]\alpha}_\gamma\,
R^\gamma_{\mu\sigma\beta}$,
$ \, \, f^{\alpha\beta}_\mu=-4/3 \nabla_\sigma
\F^{[\alpha\sigma]\beta}_\mu$
and 
\be
\U^{\alpha\beta}=\left\{ \F^{[\alpha\beta]}_\mu -2/3 \nabla_\sigma
\F^{[\alpha\beta]\sigma}_\mu\right\} \xi^\mu +4/3 \F^{[\alpha\beta]\sigma}
_\mu \nabla_\sigma \xi^\mu\label{FFTILDE2}
\ee
The generalization to $s\neq 0$ is easily obtained replacing lower Greek
indices with Latin indices, replacing the Riemann tensor with the field
strength $F$ and enlarging the covariant derivatives to act on both internal
and spacetime indices.}
\ba
\F (L',X)&=&\tilde \F (L',X)+\div \U(L',X)\label{splitting}\\
&=&\left\{
\tilde
\F^{\rho}_\mu\xi^\mu+\tilde\F^{(\rho\rho_1)}_\mu\nabla_{\rho_1}\xi^\mu+\dots
+\tilde\F^{(\rho\rho_1\dots\rho_{h
+r-1})}_\mu\nabla_{(\rho_1\dots\rho_{h
+r-1})}\xi^\mu\right.\nonumber \\
&&\left.+\tilde\F^{\rho}_A\xi^A+\tilde\F^{(\rho\rho_1)}_A\nabla_{\rho_1}\xi^A+\dots
+\tilde\F^{(\rho\rho_1\dots\rho_{h
+s-1})}_A\nabla_{(\rho_1\dots\rho_{{h
+s-1}})}\xi^A
\right\}\, ds_\rho\nonumber\\
&&+d_\sigma \U^{[\rho \sigma]}\,ds_\rho\nonumber 
\ea
Here and in the sequel square brackets around indices denote
skew--symmetrization. (For the  general setting of  the above decomposition 
for any pair $(r,s)$ we refer the reader to 
\cite{Robutti}). We only stress that this kind of integration by parts  is a
well--defined global operation  in the bundle framework. Globality is indeed
assured  by the direct use of covariant derivatives. Uniqueness  of $\tilde\F$
instead descends  from its symmetry properties and from the fact that  $\div$
is  a nilpotent operator: $\div\circ \div=0$.
Therefore,
it can  easily checked that  the term $\tilde \F (L',X)$ in the right hand
side of \form{splitting}  coincides with the variation
$\delta_X\tilde\E(L,\Xi)$ of the reduced current and it is hence vanishing
on--shell. Indeed, inserting \form{splitting} into \form{F-E} and taking  
$\div \circ \div=0$ into
account we have:
\be
\div \left\{\tilde \F (L',X)-\delta_X\tilde\E(L,\Xi)\right\}=0\label{F-tildeE}
\ee
Since  the components $\xi^\mu$ and $\xi^A$ are arbitrary and since both 
terms 
$\tilde \F (L',X)$ and $\delta_X\tilde\E(L,\Xi)$ are, by construction,
symmetric in the upper indices, it is easy to verify that \form{F-tildeE}
implies $\tilde \F (L',X)=\delta_X\tilde\E(L,\Xi)$. 

Also the potential $\U$ features an expression  of the kind:
\ba
\U^{[\rho\sigma]}&=&\U^{[\rho\sigma]}_\alpha\xi^\alpha+\dots +
\U^{[\rho\sigma](\rho_1\dots \rho_{h+r-2})}_\mu \nabla_{(\rho_1\dots
\rho_{h+r-2})}\xi^\mu\nonumber\\
&+&\U^{[\rho\sigma]}_A\xi^A+\dots +
\U^{[\rho\sigma](\rho_1\dots \rho_{h+r-2})}_A \nabla_{(\rho_1\dots
\rho_{h+s-2})}\xi^A
\ea
which  is clearly not unique  but only defined  modulo closed forms.
Nevertheless (see \cite{Robutti}) there exists a unique representative  the
coefficients  of which have the maximal symmetry property:
$\U^{[\rho\sigma\rho_1]\dots
\rho_{p}}=0$. From now on we shall select this  representative which is, from
a  mathematical viewpoint, the \emph{canonical} one. The physical viability of
this choice will be tested in applications.\vspace{.7cm}

\noindent Let us summarize. Starting from the variational Lagrangian
\form{Lprimo} we have obtained a  current:
\be
\F (L',X)=\delta_X\tilde \E (L,\Xi)+\div \U(L',X)\label{current}
\ee
which is conserved on--shell (see \form{12}) and it is also exact on--shell if
$X$ is a solution of the linearized field equations, since in this case the
term
$\delta_X\tilde \E (L,\Xi)$ in
\form{current} is vanishing (notice, instead, that the current $\F
(L',X)-\delta_X\tilde \E (L,\Xi)$ is exact also off-shell). \vspace{.5truecm}

Let us now consider a section $\varphi: M\ram Y$. Its prolongation
$j\varphi$ to a suitable (finite) order can be used to pull-back expression
\form{current} onto the base manifold.
We shall   denote by $\F ( X,\varphi)$, $\tilde \E
(\Xi,\varphi)$ and $\U (X,\varphi)$ the pull-backs on $M$ of the relevant
quantities in \form{current}  (from now  on the
indication
$L'$ between the round brackets will be omitted for the sake of simplicity).
Given a hypersurface $D$ in spacetime we  define the \emph{variation}
$\delta_X\Q_D(\Xi,  \varphi)$ of the conserved quantity $\Q_D(\Xi, 
\varphi)$, relative to the set $(\Xi,
\varphi, D)$  as follows:
\ba
\delta_X\Q_D(\Xi,  \varphi)&:= &\int_D   \F (X,\varphi)=\int_D
\delta_X\tilde
\E (\Xi,\varphi)+\int_{\partial D }
\U(X,\varphi)\label{currentint}\\
&\simeq&\int_{\partial D }
\U(X,\varphi)\label{currentint2}
\ea
where the last equality holds true  whenever  $\varphi$ is a solution of
field equations and $X$ is tangent to the space of the solutions, i.e. 
it describes a one--parameter  deformation of $\varphi$ along nearby
solutions. In particular if the region $D$ is a (portion
of a) Cauchy surface and the vector field $\xi$ is transverse to $D$ we
identify $\delta_X\Q_D(\Xi,  \varphi)$ in \form{currentint}  with the
variation of the Hamiltonian
$\delta_X\H_D(\Xi,  \varphi)$. We stress that, for the time being, this is
just a definition. Nevertheless the name  Hamiltonian  will be
justified   by  the  applications we shall explicitly consider. Indeed it will
turn out   that
$\delta_X\H_D(\Xi,  \varphi)$ physically describes the evolution of the fields
along the flow of the vector $\xi$.
\begin{Remark}{\rm We  remark  that, even though we have assumed a
gauge natural Lagrangian as the starting point for our framework,  the whole
theory is in fact developed from the Euler--Lagrange morphism \form{Lprimo}.
The Lagrangian is necessary, a priori, only in defining the dynamics via its
equations of motion. This implies that
it is of no interest whatsoever Lagrangian representative we choose in the
cohomology class
$[L]$  the elements of which differ from each other only by the addition of
divergences. Indeed all the representatives inside the class $[L]$ give rise to
the same equations. Even more, we can relax the hypothesis that the Lagrangian
be
 gauge natural since only the equations of motion are required to be
gauge invariant  as well as diffeomorphism invariant. Hence, our formalism
allows to encompass also more general classes of field theories, e.g.
Chern--Simons theories. 
\CVD
}
\end{Remark}

\begin{Remark}{\rm Notice that, through a  direct application of the Noether
uheorem, starting from the Lagrangian $L$ and from the infinitesimal generator
of symmetries
$\Xi$,  one is able to algorithmically construct a conserved Noether current 
$\E (L,\Xi)$ and to exhibit a superpotential $\V(L,\Xi)$  such that the
following two identities hold:
\ba
&&\div \E (L,\Xi)=L'(L, \Xi)\simeq 0\label{N1}\\
&&\E (L,\Xi)=\tilde \E
(L,\Xi)+\div \V(L,\Xi)\label{Noeth}
\ea
 where $\tilde \E
(L,\Xi)$ is the same reduced current appearing in \form{current} (see
\cite{Remarks,Lagrange} for details). By taking the variation of \form{N1} and
taking
\form{deltaL'} into account
 we then get:
\be
\div \delta_X \E (L,\Xi)= \div \F (L',X)
\ee
so that $\delta_X \E (L,\Xi)$ and $\F (L',X)$   differ from each other for a
form which  (at least locally) is exact:
\ba
\F (L',X)&=&\delta_X \E (L,\Xi)+\div \C(L,\Xi, X)\label{TTT}\\
&\um^{\form{Noeth}}&\delta_X \tilde \E (L,\Xi)+\div\left\{\delta_X \V(L,\Xi)+ 
\C(L,\Xi, X)\right\}\nonumber
\ea 
Notice now that the  divergence $\div \C$ cannot be vanishing in general.
Indeed, if $\div \C$  vanishes, a comparison between \form{TTT} and
 \form{current}  leads to 
  the identification  $\delta_X \V
(L,\Xi)=\U(L',X)$, which cannot be true. We can
infer this without any calculation. Indeed, it is well known that if 
we consider a  Lagrangian $\bar L=L+\div \alpha$ which differs from a
given   Lagrangian $L$ only for the addition of a divergence, then the
reduced current remains the same, i.e. $\tilde \E
(\bar L,\Xi)=\tilde \E
(L,\Xi)$ while the Noether superpotential transforms as follows (see
\cite{forth,Wald}):
\be
\V(\bar L,\Xi)=\V(L,\Xi)+i_\xi \alpha\label{ULL}
\ee
 On the other hand, being equation
\form{current} built out directly from the equations of motion \form{Lprimo},
is not affected by the addition of divergence terms to the Lagrangian. For
this reason we see that 
 the divergence term $\div \C(L,\Xi, X)$ cannot be vanishing in general and it
must 
 be sensitive to the representative we choose inside
the cohomology class $[L]$ a given Lagrangian belongs to, in order to
counterbalance  the transformation rule \form{ULL} into \form{TTT}. 

Roughly speaking, we could say that the formalism developed up to now, namely
the construction of $\div \U(L',X)$ in \form{current}, is nothing but a recipe
to algorithmically define the additional term $\div \C(L,\Xi, X)$ which has to
be added to the Noether superpotential in order to make the variation of
conserved quantities insensitive to the choice of a representative inside
$[L]$. 

When dealing with applications we shall see  that the fibered morphisms
$\U(L',X)$ and 
$\delta_X \V (L,\Xi)$ differ indeed for a term which is nothing but the
\emph{covariant} Regge--Teitelboim boundary correction term; see
\cite{Remarks,CADM,forth,Silva,RT}.\CVD }
\end{Remark}
\vspace{.5truecm}

\section{Purely metric formulation of gravity}
\label{section4}
Let us now consider the Hilbert Lagrangian density
\be
\L_H={1\over 2k}\sqrt{g} g^{\mu\nu} R_{\mu\nu}(j^2g) \label{19}
\ee
($k=8\pi G/c^4$)
or any other Lagrangian differing from it by the addition of boundary terms 
(e.g. the first order Einstein Lagrangian \cite{EinsteinL}, the first order
covariant Lagrangian \cite{BTZ,by,CADM,Katz} or the York's Trace--K
Lagrangian
\cite{BY,York}). Indeed,  what is really relevant is only  the expression of  
 Einstein's field  equations:
\be
-{\sqrt{g}\over 2k}\, G^{\mu\nu} =-{\sqrt{g}\over
2k}\,\left\{R^{\mu\nu}-{1\over 2} g^{\mu\nu}\,R\right\}=0\label{20}
\ee
from which we define the variational Lagrangian \form{Lprimo}:
\be
L'(\xi)= {\sqrt{g}\over 2k}\, G^{\mu\nu}\,\pounds_\xi g_{\mu\nu}\, ds
={\sqrt{g}\over k}\, G^{\mu\nu}\,\nabla_\mu \xi_\nu\, ds\label{LprimoGR}
\ee
where $\xi=\xi^\mu\partial_\mu$ denotes  a vector field in
spacetime.\vspace{.7cm}

\noindent {\bf First step.} Let us consider a vertical vector field
$X=X_{\mu\nu}{\partial
\over
\partial g_{\mu\nu}}=\delta g_{\mu\nu}{\partial \over \partial g_{\mu\nu}}$ on
the configuration bundle $Y={\mathop{\rm Lor(M)}\nolimits}$ of Lorentzian
metrics and let us perform the variation $\delta_X L'(\xi)$. By taking into
account the relations:
\ba
&&\delta R_{\mu\nu}=\nabla_\rho (\delta u^\rho_{\mu\nu}), \qquad
u^\rho_{\mu\nu}=\gamma^\rho_{\mu\nu}-\delta^\rho_{(\mu}
\gamma^\sigma_{\nu)\sigma}\\ &&\delta
\gamma^\rho_{\mu\nu}={1\over2}g^{\rho\sigma} ( -\nabla_\sigma \delta g_{\mu\nu}
+\nabla_\mu \delta g_{\nu\sigma}
+\nabla_\nu \delta g_{\sigma\mu})
\ea
(where $\gamma^\rho_{\mu\nu}$ denotes the Levi--Civita connection of the
metric $g$) we obtain: 
\be
\delta_X L'(\xi)=d_\gamma \F^\gamma(j^2g, j^1 X,j^2 \xi)\, ds
\ee
where:
\be
\F^\gamma=\F^\gamma_\rho (j^2g,X) \xi^\rho+
\F^{\gamma\lambda}_\rho (j^1g,j^1X) \nabla_\lambda \xi^\rho+
\F^{\gamma(\lambda\sigma)}_{\rho} (g,X) \nabla_{(\lambda\sigma)} \xi^\rho
\label{24}\ee
with:
\ba
\F^\gamma_\rho&=&{\sqrt{g}\over 2k} G^{\alpha\beta}\,\delta g_{\alpha\beta}
\,\delta_\rho^\gamma +{1\over 2}T^{\gamma[\mu\lambda]}_\sigma
R^\sigma_{\rho\mu\lambda}
\nonumber \\
\F^{\gamma\lambda}_\rho&=&
{\sqrt{g}\over 2k}\,\delta u^\gamma_{\mu\nu}(2 g^{\lambda\mu} \,
\delta^\nu_\rho-g^{\mu\nu}\, \delta ^\lambda_\rho)
\\
\F^{\gamma(\lambda\sigma)}_{\rho}&=&T^{\gamma(\lambda\sigma)}_\rho
\nonumber\\
T^{\gamma\rho\lambda}_\sigma&=&{\sqrt{g}\over k} \,\delta g_{\alpha\beta}
\left(g^{\beta[\lambda}g^{\rho]\gamma}\delta^\alpha_\sigma+
g^{\rho\beta}g^{\alpha[\gamma}\delta^{\lambda]}_\sigma\right.\nonumber\\
&&\left.+
{1\over2} g^{\alpha\beta}(g^{\rho[\lambda}\delta^{\gamma]}_\sigma+
g^{\gamma[\lambda}\delta^{\rho]}_\sigma)
\right)
\ea
\noindent {\bf Second step.} Integrating  the expression \form{24} by parts
according to the formulae \form{FFTILDE1} and  \form{FFTILDE2} we obtain
\be
\F^\gamma=\delta_X\tilde \E^\gamma(L,\xi)+d_\rho \U^{[\gamma \rho]}
\label{26}\ee 
where:
\ba
&&\tilde \E^\gamma(L,\xi)={\sqrt{g}\over k} G^{\gamma}_{ \nu} \xi^\nu\\
&&\U^{[\gamma \rho]}=\delta \left\{ {\sqrt{g}\over k}\nabla^{[\rho}
\xi^{\gamma]}\right\} + {\sqrt{g}\over k}\, g^{\mu\nu}\,\delta
u^{[\rho}_{\mu\nu}\,\xi^{\gamma ]}\label{30}
\ea
Notice that the reduced current  $\tilde \E(L,\xi)$ is clearly vanishing 
on--shell. Moreover the first term in the right hand side of \form{30} is
nothing but 
 the
variation of the Noether superpotential (i.e., in our case,   the variation of
the Komar superpotential; \cite{Komar}).
 On the contrary the second term  in the right
hand side of \form{30} is \emph{the covariant Regge--Teitelboim correction
term} \cite{Booth,CADM,forth}, namely it is the term which has to be
added to the Hamiltonian  to define a well--posed variational principle. To
understand this latter statement,  let us  analyse \form{26} in details. 
Let us choose a (local) section  $g: x\mapsto g_{\mu\nu}(x)$ of the
configuration bundle and let us make use of it to pull--back formula \form{26}
onto spacetime $M$. Let  $D$ be a  (portion
of a) Cauchy surface in $M$ and let the vector field $\xi$ be transverse to
$D$. According with our previous definition  we identify the integral 
(of the pull--back) of \form{26} with the variation of the Hamiltonian:
\ba
\delta_X\H_D(\xi,  g)&=&\int_D \delta\left\{{\sqrt{g}\over k} G^{\gamma}_{
\nu}
\xi^\nu \right\}\,ds_\gamma\label{29}\\
&+&\int_{\partial D}\left\{\delta \left [{\sqrt{g}\over 2k}\nabla^{[\rho}
\xi^{\gamma]}\right] + {\sqrt{g}\over 2k}\, g^{\mu\nu}\,\delta
u^{[\rho}_{\mu\nu}\,\xi^{\gamma ]}\right\} ds_{\rho\gamma}\nonumber
\ea
Since the vector field $\xi$ is transverse to $D$ we can drag (at least
locally) the surface $D$ along the flow of $\xi$, defining in this way a
$m$--dimensional region of spacetime which is foliated into hypersurfaces
diffeomorphic to $D$. By making use of the  (3+1) ADM formalism we can rewrite
\form{29} in terms of quantities which are adapted to the foliation (the
details of such a calculation can be found in \cite{Booth,BLY,forth}). We
obtain:
\be
\delta_X\H_D(\xi,  g)=\int_{D}\left\{\delta N
\H +\delta N^\al\H_\al +[h_{\al\b}] \delta P^{\al\b}-[P^{\al\b}]
\delta h_{\al\b}\right\}d^3x\label{30000}
\ee
where $N$ and  $N^\al$ are, respectively, the lapse and the shift  of the
vector field $\xi=\partial_t$, $\H$ and $\H_\al$ are the usual
Hamiltonian constraints of General Relativity, $P^{\al\b}$ is the momentum
conjugated to the metric $h_{\al\b}$ on the hypersurface $D$ and 
$[h_{\al\b}]$ and $\left[P^{\al\b}\right]$ denote the right hand side of the
Hamilton equations: $\pounds_\xi h_{\al\b}=[h_{\al\b}]$ and $\pounds_\xi
P^{\al\b}=\left[P^{\al\b}\right]$ (see
\cite{Booth,Gravitation}).
We  point out 
 that the boundary terms  arising  in the variation of the
first term  in the right hand side of
\form{29} exactly cancel out  the  second and the third 
  boundary terms. Hence
 we end up with the ``pure''  bulk term \form{30000} which is the
correct expression one would expect for the  variation of the  Hamiltonian for
General Relativity; see \cite{Booth,Boothultimo,BLY,forth,RT}. This fact
justifies, a posteriori, the definition of variation of the Hamiltonian we
have previously attributed to formula \form{29}.

We just outline that the \emph{variation} of the energy  $\delta_X E_D(\xi, 
g)$ enclosed in the quasilocal region bounded by $\partial D$ and relative to
the pair $(\xi, g)$ can be defined as  the on--shell value of
\form{29}, namely:
\be
\delta_X E_D(\xi,  g)=
\int_{\partial D}\left\{\delta \left [ {\sqrt{g}\over 2k}\nabla^{[\rho}
\xi^{\gamma]}\right] + {\sqrt{g}\over 2k}\, g^{\mu\nu}\,\delta
u^{[\rho}_{\mu\nu}\,\xi^{\lambda ]}\right\} ds_{\rho\gamma}
\label{31}
\ee
where, we recall, for  \emph{variation} we mean the infinitesimal change
of energy when we move along nearby solutions (different vectors $X$
corresponding to different paths in the space of solutions all passing 
through  the same ``point'' $g(x)$). The (3+1) splitting of formula
\form{31} can be found in \cite{forth} and for this reason  we do not exhibit
it here. In
\cite{Anco,Barnich,Nester,by,forth,Silva} 
 the relations interplaying between the integrability conditions of \form{31}
and   boundary conditions on the metric and its derivatives  are analysed in
 detail. We just point out that expression
\form{31}  with Dirichlet boundary conditions gives rise to the Brown--York
quasilocal energy \cite{by,forth}. Moreover the implications  of \form{31} in
relation with the first law of black hole thermodynamics  have been discussed
in
\cite{HIO,Ultimo,BY,ML,Remarks,Taub,forth,Wald}. It was there also emphasized
how this formula  sheds some light on the fundamental contribution of geometry
in characterizing  the entropy  of black hole solutions and even of more
general solutions of Einstein's equations.

 For all
these reasons we consider expression \form{31} as an acceptable and 
viable definition for the   (variation of) energy.

\section{Gravity in tetrad formalism}
\label{section5}
Let us now consider the formulation of gravity in the tetrad formalism. Even if
the  tetrad formulation  is in many cases equivalent to the purely metric
formalism,  it becomes necessary when dealing with spinor matter, so that it
deserves an investigation  in its own.

It is well known that a
good  mathemathical arena to globally describe tetrad gravity is the gauge
natural bundle framework (see \cite{{Lorenzo,Godina,Matteucci}}). In this
context the structure bundle of the theory is a principal bundle $(P, M, p;
SO(1,3))$  over spacetime
$M$.\footnote{Since we shall not consider here the explicit coupling between
gravity and spinor matter the structure group is assumed to be the pseudo
orthogonal group instead of its  twofold covering group $Spin(1,3)$} According
to \cite{Lorenzo} a tetrad field is defined to be a section of a $GL(4,\Re)$ 
bundle
$\Sigma$ which is the  bundle associated to  the
 bundle $P\times L(M)$, where  $L(M)$ denotes the frame bundle,  via the left 
action 
\ba
\rho&:&
(SO(1,3)\times GL(4,\Re))\times GL(4,\Re)\ram  GL(4,\Re)\nonumber\\
\rho&:&(\Lambda, J; X)\mapsto
\Lambda\cdot X\cdot (J^{-1})
\ea
Fibered coordinates on $\Sigma$ are denoted by $(x^\mu, e^i_\mu)$, where
$i=0,\dots 3$. A tetrad field is then a map
$x\mapsto (x^\mu, e^i_\mu=\theta^i_\mu(x))$. Given  a tetrad field we can
define a metric over spacetime through the rule $g_{\mu\nu}= \eta_{ij} 
\theta^i_\mu \theta^j_\nu$, $\eta=diag(-1,1,1,1)$,  so that, a posteriori, the
structure bundle $(P, M, p; SO(1,3))$ can be identified  with the bundle of
orthonormal frames $SO(M,g)$ which is the subbundle of the frame bundle  $L(M)$
formed by all $g$--orthonormal  frames. We stress that, in this framework, the
index
$i$ of the tetrad is not merely a label denoting a set of four $1$--forms in
spacetime. In this case tetrad fields  would be  merely a \emph{local} basis 
of the tangent space (globality  being recovered  only for parallelizable 
manifolds)  and 
the theory would just be a natural theory. Global frames, i.e. tetrads, are
instead truly gauge natural objects: they are  sensitive to the
transformations  of the structure bundle $P$. Each  authomorphism of the
principal bundle functorially induces, via 
the left action $\rho$, a transformation law acting on the tetrad. This
transformation law allows to canonically define the Lie derivative of the
tetrad fields with respect to any vector field on the structure bundle.
Indeed    a projectable vector field $\Xi_P$ in the principal bundle, locally
described as 
\be\Xi_P= \xi^\mu(x)\partial_\mu + \xi^{ij}(x)
\rho_{ij},\qquad\xi^{ij}=\xi^{[ij]}\label{35}
\ee
 (having denoted by $\rho=g\,\partial/\partial g$, $g\in SO(1,3)$
a basis  for right invariant vector fields on $P$ in a trivialization  $(x,g)$
of $P$) canonically induces the projectable vector field $\Xi$ on $\Sigma$
given by
$\Xi=\xi^\mu\partial_\mu +\xi^i_j\, e^j_\mu \partial/ \partial
e^i_\mu$. Through  the vector field $\Xi$ we can define the  formal Lie
derivative
$\pounds_\Xi e:J^1\Sigma\ram V(\Sigma)$. It  is intrinsically defined
by the rule $\pounds_\Xi e\circ j^1\theta :=\pounds_\Xi\theta=T\theta\circ
\xi-\Xi\circ \theta$, where $\xi$ denotes the projection of $\Xi$ onto $M$ and
$\theta $ is a tetrad field. In coordinates:
\be
\pounds_\Xi e^i_\mu=\xi^\nu d_\nu e^i_\mu+d_\mu \xi^\nu  e^i_\nu -\xi^i{}_j
e^j_\mu\label{Lietetrade}
\ee
In the tetrad affine formulation of gravity the
configuration bundle $Y$ is assumed to be   the  bundle:
$
Y=\Sigma\ram M
$. 
 The Lagrangian of the theory turns out to be
the  fibered morphism
$
L: J^2\Sigma\ram \Lambda^4(M)
$
locally described through the Lagrangian density:
\be
\L={1\over 8k} e^i_\mu\, e^j_\nu\, R^{kl}_{\alpha\beta}(j^2 e)\,
\epsilon^{\mu\nu\alpha\beta} \epsilon_{ijkl} \label{Ltetrade1}
\ee
where
$ R^{kl}(j^2e)=d \omega^{kl}+\omega^k{}_h\wedge \omega^{hl}$ is the field
strength of the Levi--Civita spin  connection  $\omega^{kl}$. The theory is
invariant under the whole group
$\Aut(P)$ meaning that each vector field \form{35} is an infinitesimal
generator of symmetries. Field equations ensuing from \form{Ltetrade1} are,
\be
{1\over 4k} e^j_\nu R^{kl}_{\alpha\beta}
\epsilon^{\mu\nu\alpha\beta}\epsilon_{ijkl}=0\label{39AA}
\ee
As it is well known  the theory described
by
\form{39AA} is completely  equivalent to Einstein's theory (provided that
$\det(e^i_\mu)\neq 0$), the difference between the two sets of field equations 
being only a matter of notation. Therefore, we should expect to reproduce the
same expressions for the conserved quantities we derived in the previous
section. 

The variational Lagrangian associated with the Euler--Lagrange
equations 
\form{39AA} and the infinitesimal generator of symmetries \form{35}
turns out to be:
\be
L'(\Xi)=-{1\over 4k}
\epsilon^{\mu\nu\alpha\beta}\epsilon_{ijkl}
\left\{ e^j_\nu R^{kl}_{\alpha\beta}\pounds_\Xi  e^i_\mu
\right\}ds\label{4111}
\ee
where the Lie derivative of the tetrad field has been defined by
\form{Lietetrade}. It is an easy calculation to show that \form{4111} coincides
with \form{LprimoGR} so that exactly the same results of the previous section
are recovered. 
Notice that, even though this is exactly the result we physically  expected, it
it  nevertheless a result not so obvious from a mathematical viewpoint. 
Indeed, even if the Lagrangians \form{19} and \form{Ltetrade1} 
are equivalent as far as  their field equations contents are concerned, they
describe instead  different theories from a geometrical viewpoint, since the
former theory is a natural theory while  the latter one is a true gauge natural
theory. Therefore they admit two different groups of symmetries given,
respectively, by $\Diff(M)$ and $\Aut(P)$. For this reason conserved quantities
for tetrad gravity calculated via Noether theorem  are
 \emph{intrinsically indeterminate}  from a mathematical viewpoint (see
\cite{Matteucci}) since the components  $\xi^{ij}$ of the vector field
\form{35} enter into the Noether superpotential and they can be, a priori, 
  fixed arbitrarily. This indeterminacy does not occur in our framework since,
in
\form{4111},  the skew--symmetric components $ \xi^{ij}$ are contracted with
the symmetric  components $R_{\mu\nu}e_i^\mu e_j^\nu$ of the Ricci tensor and
therefore they give rise to an identically vanishing term. 

Therefore, we expect that in any 
generalization of Einstein's theory involving also the non--symmetric part of
the Ricci tensor (see, e.g. \cite{FK,Hehl}) the indeterminacy will still remain
so that we have to somehow  bypass it. This is the issue we shall consider in
the next section.

\section{Einstein Cartan theory and the Kosmann lift}\label{KosmannS}
The Einstein--Cartan (Sciama--Kibble) theory, ECKS from now on,  describes a
generalization of Einstein theory  of gravitation in which the spin  angular
momentum  of matter plays a dynamical role. In the ECKS model  the
dynamical connection is still metric compatible  but it is no longer symmetric,
the torsion part being  coupled  to the spin of matter. Therefore the theory
leads  to deviations  from General Relativity (even though  the mathematical
differences can be physically  appreciated only in extreme  solutions);
see \cite{Hehl}.

The  ECKS theory can be framed in the domain   of   gauge natural theories;
see \cite{Matteucci}.  The configuration bundle
$Y$ for the  gravitational ECKS Lagrangian is assumed to be   the product
bundle:
\be
Y=\Sigma\times \C(P)\ram M
\ee 
where $\C(P)=J^1P/SO(1,3)$ denotes the bundle of principal connections over the
structure bundle $P$ (see \cite{Kolar}) while $\Sigma$ and $P$  have been
defined in the previous section. The bundle
$Y$ is clearly a gauge natural bundle associated with the structure bundle
$(P, M, p; SO(1,3))$. Fibered coordinates on
$Y$ are denoted by
$(x^\rho, e^i_\mu, A^{jk}_\nu)$. The Lagrangian of the theory is the sum 
$L=L_{EC}+ L_M$ of the gravitational ECKS Lagrangian $L_{EC}$ and the matter
Lagrangian $L_M$. The former Lagrangian   turns out to be the  fibered morphism
$
L_{EC}: \Sigma\times J^1\C(P)\ram \Lambda^4(M)
$
locally described through the Lagrangian density:
\be
\L_{EC}={1\over 4} e^i_\mu\, e^j_\nu\, F^{kl}_{\alpha\beta}(j^1A)\,
\epsilon^{\mu\nu\alpha\beta} \epsilon_{ijkl} \label{Ltetrade}
\ee
where
$ F^{kl}(j^1A)=d A^{kl}+A^k{}_h\wedge A^{hl}$ is the field strength of the
connection (from now on we shall omit the constant factor $1/2k$ in front of
the Lagrangian).  The theory is invariant under the whole group
$\Aut(P)$ (provided that the matter Lagrangian is, in its turn, a gauge
natural Lagrangian) meaning that each vector field
\form{35} is an infinitesimal generator of symmetries. Field equations
ensuing from
\form{Ltetrade} through the variations of the independent fields $e^i_\mu$ and
$A^{jk}_\nu$ are, respectively:
\ba
&&{1\over 2} e^j_\nu F^{kl}_{\alpha\beta}
\epsilon^{\mu\nu\alpha\beta}\epsilon_{ijkl}=\tau_i^\mu\label{39A}\\
&&-\epsilon^{\mu\nu\alpha\beta}\epsilon_{ijkl} e^i_\mu \DA_\alpha e^j_\nu
=\tau_{kl}^\beta\label{39B}
\ea
where $\DA_\alpha e^i_\mu=d_\alpha e^i_\mu+A^i{}_{j\alpha} e^j_\mu$, while
$\tau_i^\mu$ and $\tau_{kl}^\beta$ are, respectively,  the energy momentum
and the spin  of matter, the explicit form of which depends on  the matter
Lagrangian.

As it is well known, if the spin of matter is zero the theory described
by
\form{39A} and
\form{39B} is on--shell equivalent to Einstein's theory (provided that $\det
(e^i_\mu)\neq 0$). Indeed equation
\form{39B} admits the solution $A^i{}_{j\alpha}=\omega^i{}_{j\alpha}(j^1 e)$,
where $\omega^i{}_{j\alpha}(j^1 e)$ denotes the  Levi--Civita
connection built out of the tetrad together with  its derivatives. This
solution inserted back into the Lagrangian
\form{Ltetrade} gives rise to the Hilbert Lagrangian \form{19} while equations
\form{39A} become Einstein's equations \form{20}. Nevertheless, even when
$\tau_{kl}^\beta=0$ the equivalence holds true only on--shell since there
does not exist any off--shell relation between the tetrad field and the
connection. On the contrary,  when the spin
of matter does not vanish equation \form{39B} becomes an algebraic
equation relating the ``non--metricity''  of the connection (namely, the
torsion) with the spin matter content. 

We shall now exhamine the energetic informations ensuing from  the pure 
gravitational part of the theory (left hand side of equations \form{39A} and 
\form{39B}). Since all calculations  will be performed off--shell, the result
will be not affected by the specific distribution of spinning matter one is
taking into consideration. The variational Lagrangian associated with   the
left hand side of 
Euler--Lagrange equations 
\form{39A}--\form{39B} and the infinitesimal generator of symmetries \form{35}
turns out to be:
\be
L'(\Xi)=
\epsilon^{\mu\nu\alpha\beta}\epsilon_{ijkl}
\left\{-{1\over 2} e^j_\nu F^{kl}_{\alpha\beta}\pounds_\Xi  e^i_\mu
+ e^i_\mu \DA_\alpha e^j_\nu \pounds_\Xi A^{kl}_\beta\right\}ds\label{41}
\ee
where the Lie derivative of the tetrad field has been defined in
\form{Lietetrade}, while 
\ba
\pounds_\Xi A^{kl}_\beta&=&\xi^\rho d_\rho  A^{kl}_\beta +d_\beta \xi^\rho 
A^{kl}_\rho+\DA_\beta \xi^{kl}\nonumber\\
&=&\xi^\rho F^{kl}_{\rho\beta}+\DA_\beta \xi^{kl}_{(V)}\qquad \qquad
(\xi^{kl}_{(V)}=\xi^{kl}+ A^{kl}_\rho\xi^\rho)\label{43}
\ea
Notice that the Lagrangian \form{41} is a gauge natural
Lagrangian by its own; see section  \ref{Natural and Gauge Natural Theories}.

{\bf First step.} The 
variation of 
\form{41} with respect to a vertical vector field  $X=\delta e^i_\mu
\partial/\partial e^i_\mu+ \delta A^{ij}{}_\mu\partial/\partial A^{ij}{}_\mu$
splits into  a term 
$<e(L')\vert X>$ which is identically vanishing plus a divergence term $\div
\F(L',X)$ (see \form{deltaL'}).  The term  $\F(L',X)$ is linear in the
coefficients
$\xi^\mu,
\xi^{ij}$ and their  derivatives $d_\mu \xi^\nu, \DA_\mu
\xi^{ij}$. Integrating by parts according to formula \form{FFTILDE1} and
\form{FFTILDE2}  we get ({\bf second step}):
\be
\F^\gamma (L',X)=\delta_X\tilde \E^\gamma (L,\Xi)+d_\beta
\U^{\gamma\beta}(L',X)
\ee 
where:
\ba
&&\tilde \E^\gamma (L,\Xi)=\epsilon^{\gamma\nu\alpha\beta}\epsilon_{ijkl}
\left\{-{1\over 2} e^j_\nu F^{kl}_{\alpha\beta} e^i_\rho\xi^\rho
-e^i_\beta \DA_\alpha e^j_\nu\, \xi^{kl}_{(V)}\right\}\\
&&\U^{\gamma\beta}(L',X)=\epsilon^{\gamma\beta\mu\nu}\epsilon_{ijkl}
e^j_\nu\left\{ \delta A^{kl}{}_\mu  e^i_\rho\xi^\rho+\delta e^i_\mu
\xi^{kl}_{(V)}\right\}\label{cumber}
\ea
Therefore, the variation of conserved quantities is defined (see
\form{currentint}) as the on--shell
integral:
\ba
\delta_X\Q_D(\Xi, e,A)&\simeq&{1\over 2}\int_{\partial D
}\U^{\gamma\beta}(L',X)ds_{\gamma\beta}\nonumber\\
&\simeq&{1\over 2}\int_{\partial D
}\epsilon^{\gamma\beta\mu\nu}\epsilon_{ijkl}
e^j_\nu\left\{ \delta A^{kl}{}_\mu  e^i_\rho\xi^\rho+\delta e^i_\mu
\xi^{kl}_{(V)}\right\}ds_{\gamma\beta}\label{indet}
\ea
where $\partial D$ is the two dimensional boundary of a three dimensional
region
$D$.
Notice that the components $\xi^{ij}$ of the vector field enter into the
definition of conserved quantities. Since, up to now,  no preferred condition
can be mathematically  imposed  on those components, formula
\form{indet} features an intrinsic indeterminacy. Nevertheless it was
suggested in \cite{Matteucci}  that 
   the indeterminacy can be eliminated if the vertical components
$\xi^{kl}_{(V)}$ are defined through the Kosmann lift (we shall enter into 
details later on). The choice of the
Kosmann lift  was  justified, a posteriori, in \cite{Matteucci} from a
 physical viewpoint, since, in an  apparently surprising way, it is the only
lift among the possible ones which  allows one to exactly reproduce the
expected values for conserved quantities in explicit applications; see
\cite{Ultimo,BTZ}. Our next issue will be then to give a mathematical 
justification for this choice. Indeed we shall show  that  the Kosmann lift 
arises spontaneously  when trying to restore the naturality of the gauge
natural variational Lagrangian
\form{41}.

To this end let us  face up to the problem
of calculating, through the variational Lagrangian
\form{41}, the variation of the conserved quantity associated to a spacetime
vector field $\xi$. Notice, however, that
 vertical vector fields $\Xi_P=\xi^{ij}(x)
\rho_{ij}$   in the structure bundle (see
\form{35}) are well defined objects 
and they describe infinitesimal gauge transformations (and, accordingly, 
expression
\form{indet} describes the variation of gauge charges when calculated for
vertical vector fields).
 On the contrary,   horizontal vector
fields $\Xi_P= \xi^\mu(x)\partial_\mu$  are not globally well defined
since they do not transform tensorially.
 Roughly
speaking this means  that the Lie derivatives of the dynamical fields are
defined only with respect to vector fields in $P$ and \emph{not} with respect
to spacetime vector fields. Hence,  
  we have, first of all,  to somehow lift the
vector field
$\xi$ to a vector field
$\Xi=\hat \xi$ on the principal bundle $P$. If
we are able to do this, the Lie derivatives $\pounds_{\hat \xi} A$ and
$\pounds_{\hat \xi} e$ in expression \form{41}
are well defined and 
expression \form{indet}, with $\Xi=\hat \xi$,  can be used to define
  global charges (such as energy, momentum and angular
momentum).
 What we shall   look for is, if it exists, a geometrically
well defined lift which restores, off--shell, the naturality of the variational
Lagrangian
\form{41}.  

Notice that, as far as  expression \form{41} is concerned  there 
does not appear  any preferred dynamical  linear connection on   spacetime
(as it stands, expression \form{41} already transforms covariantly with respect
to the automorphisms of the principal bundle). Hence we are free to introduce,
by hands and according to our convenience, a linear connection
$\Gamma$ built out of the dynamical fields. In this way we can extend the
$SO(1,3)$ covariant derivative $\DA$, which acts on internal (Latin) indices,
to a
$SO(1,3)\times GL(4)$ covariant derivative $\DG$ which acts on internal as well
as spacetime (Greek) indices. We define the linear connection $\Gamma$ by
requiring that $\DG$ annihilates the tetrad field, i.e.:
\be
\DG_\gamma e^i_\mu:=\DA_\gamma e^i_\mu + \Gamma^\rho{}_{\mu\gamma}\, e^i_\rho=
 d_\gamma e^i_\mu+ \Gamma^\rho{}_{\mu\gamma}\, e^i_\rho +
A^i{}_{j\gamma} \, e^j_\mu=0\label{connection}
\ee
These are  $64$ independent linear equations in the  $64$ variables
$\Gamma^\rho{}_{\mu\gamma}$. Thereby the coefficients of the linear connection,
being uniquely determined in terms of the principal connection and of
the tetrad field together with its derivatives, describe in fact a dynamical
field. Notice that from \form{connection} it immediately follows that
$g$ is parallel along $\Gamma$, i.e.:
\be
\DG_\gamma g_{\mu\nu}=0
\ee
Hence the connection turns out to be:
\be
\Gamma^\rho{}_{\mu\gamma}=\gamma^\rho{}_{\mu\gamma}(j^1g)
+K^\rho{}_{\mu\gamma}\label{gammacomp}
\ee
where $\gamma^\rho{}_{\mu\gamma}(j^1g)$ denotes the Levi--Civita connection of
the metric $g_{\mu\nu}=\eta_{ij} e^i_\mu e^j_\nu$ while $K^\rho{}_{\mu\gamma}$
is the so--called \emph{contorsion tensor}:
\be
K^\rho{}_{\mu\gamma}=T^\rho{}_{\mu\gamma}-T_\mu{}^\rho{}_\gamma-
T_\gamma{}^\rho{}_\mu, \qquad
T^\rho{}_{\mu\gamma}=\Gamma^\rho{}_{[\mu\gamma]}\label{contorsion}
\ee
(indices are clearly raised and lowered with the metric $g_{\mu\nu}$ and its
inverse).  In this step we can replace the original $24$ variables $
A^i{}_{j\gamma}$  with the $24$ variables $K^\rho{}_{\mu\gamma}$ (notice that
$K_{\rho\mu\gamma}=-K_{\mu\rho\gamma}$). 

Equation \form{connection} allows to
produce a relation between the field strength $F^{ij}{}_{\mu\nu}$ and the
Riemann tensor
\be
R^\alpha{}_{\beta\mu\nu}=d_\mu \Gamma^\alpha{}_{\beta\nu}-d_\nu
\Gamma^\alpha{}_{\beta\mu}+\Gamma^\alpha{}_{\sigma\mu}
\Gamma^\sigma{}_{\beta\nu}-\Gamma^\alpha{}_{\sigma\nu}
\Gamma^\sigma{}_{\beta\mu}
\ee
The relation is indeed established by the property:
\ba
0=[\DG_\mu, \DG_\nu]
e^i_\rho&=&F^{i}{}_{j\mu\nu}e^j_\rho-R^\gamma{}_{\rho\mu\nu}e^i_\gamma+2
T^\gamma{}_{\mu\nu}\DG_\gamma e^i_\rho=\nonumber\\
&=&F^{i}{}_{j\mu\nu}e^j_\rho-R^\gamma{}_{\rho\mu\nu}e^i_\gamma
\ea
which can be easily solved in terms of the field strength as follows:
\be
F^{i}{}_{j\mu\nu}=R^\gamma{}_{\rho\mu\nu}\, e^i_\gamma\, e^\rho_j\label{RF}
\ee
Inserting \form{connection} and \form{RF} into the variational Lagrangian
\form{41} we obtain:
\be
L'(\Xi)=\left\{2\sqrt{g}\,  (R^\mu{}_\rho - {1\over 2} \delta^\mu_\rho \, R
)\, e^\rho_i \pounds_\Xi  e^i_\mu+
\epsilon^{\mu\nu\alpha\beta}\epsilon_{ijkl} e^i_\mu\, e^j_\rho \,
T^\rho{}_{\nu\alpha}\,\pounds_\Xi A^{kl}_\beta\right\} ds\label{51}
\ee 
where $R_{\mu\nu}=R^\alpha{}_{\mu\alpha\nu}$ denotes the Ricci tensor built
out of the connection $\Gamma$. From \form{gammacomp} it follows that:
\be
R_{\mu\nu}(j^1\Gamma)=R_{\mu\nu}(j^1\gamma)+\Dg_\beta
K^\beta{}_{\mu\nu}-\Dg_\nu K^\beta{}_{\mu\beta}+ K^\beta{}_{\sigma\beta}
K^\sigma{}_{\mu\nu}- K^\alpha{}_{\sigma\nu}
K^\sigma{}_{\mu\alpha}\label{RRicci}
\ee
where $R_{\mu\nu}(j^1\gamma)$ is the Ricci tensor of the Levi--Civita
connection and $\Dg$ is the  (metric) covariant derivative induced by $\gamma$.
Notice that  $R_{\mu\nu}$ is not symmetric. By splitting the first term in the
right hand side of \form{51} into its symmetric and skew--symmetric parts  we
obtain:
\ba
L'(\Xi)&=&2\sqrt{g} \left\{G^{(\mu\rho)} e_{\rho i} \pounds_\Xi  e^i_\mu+
\delta^{\nu\alpha\beta}_{\rho\gamma\lambda} e^\gamma_k\, e^\lambda_l \,
T^\rho{}_{\nu\alpha}\,\pounds_\Xi A^{kl}_\beta\right\}\, ds\nonumber\\
&&+2\sqrt{g} R^{[\mu\rho]} 
 e_{\rho i} \pounds_\Xi  e^i_\mu\,\, ds
\label{53}
\ea
with $G^{\mu\rho}=R^{\mu\rho}-1/2 g^{\mu\rho} R$.
The above expression suggests how to select a preferred lift $\Xi$ of the
spacetime vector field $\xi$, namely, how to define the components $\xi^i_j$
in \form{Lietetrade} in terms of the the components $\xi^\mu$. We simply
require that the latter term in \form{53} does vanish, i.e:
\be
 e_{i[\rho } \pounds_\Xi  e^i_{\mu]}
=0\label{54}
\ee
This algebric equation can be easily and uniquely solved in terms of the
components
$\xi^i_j$ yielding:
\be
\xi^{ij}=\xi^{[ij]}=e^{[i}_\alpha \, e^{j]\lambda} d_\lambda
\xi^\alpha-\xi^\gamma e^{[i}_\mu d_\gamma e^{j]\mu}\label{Kosmann}
\ee
The above expression is known as \emph{the generalized Kosmann lift}.
We remark that the Kosmann lift was defined for the first time in
\cite{Lorenzo} in order to establish  a relationship  between the ad hoc
definition of Lie derivative of spinor fields given in \cite{Kosmann}  and the
general theory  of Lie derivatives on fiber bundles. We stress that, in our
framework, the generalized Kosmann lift  arises, in a completely independent
way, when trying to restore naturality  in the variational Lagrangian
\form{41}. Namely, it is the consequence of the algebraic equation
\form{54} and it is not an
\emph{ad hoc} definition.
We also stress  that the solution \form{Kosmann}  
is globally well defined. Indeed, the Kosmann lift $K(\xi)$
of a spacetime vector field $\xi$, which in   a system of fibered coordinates
read as follows:
\be
K(\xi)=  \xi^\mu\partial_\mu + \left\{e^{[i}_\alpha \, e^{j]\lambda} d_\lambda
\xi^\alpha-\xi^\gamma e^{[i}_\mu d_\gamma e^{j]\mu}\right\}
\rho_{ij}\label{KXI}
\ee
 transforms tensorially (as can be inferred through a direct
inspection) and can thereby globally defined. Namely, all the local expressions
\form{KXI} can be patched together  to define a global vector field.
\footnote{The geometric nature of the Kosmann lift can be easily understood
as follows. One starts with the spacetime vector field $\xi$ and considers its
natutal lift $L(\xi)$  to the frame bundle $L(M)$. Fixing a frame $e=(e^i_\mu)$
and thereby a spacetime metric  $g$ one can construct the bundle of
orthonormal frames $SO(M,g)$ which is a subbundle of $L(M)$. The Kosmann lift
is nothing but the projection of $L(\xi)$ from the tangent space of $L(M)$ onto
the tangent space of $SO(M,g)$; see \cite{Godina}. }

We now analyze the consequences that the
expression
\form{Kosmann} has on our theory. Inserting \form{Kosmann} into 
 \form{Lietetrade} we have:
\be
\pounds_{K(\xi)}  e^i_{\mu}={1\over 2} e^{i\nu} \pounds_\xi
g_{\mu\nu}\label{56}
\ee
 Instead if we insert \form{Kosmann} into
\form{43} and we take the relations \form{connection} and
\form{RF} into account    we get:
\ba
\pounds_{K(\xi)} A^{jk}_\mu&=&e^{[j}_\lambda\, e^{k]\gamma}\left( \xi^\sigma\,
R^\lambda{}_{\gamma\sigma\mu}+\DG_\mu \tilde \nabla_\gamma
\xi^\lambda\right)\nonumber\\
&=&e^{[j}_\lambda\, e^{k]\gamma}\pounds_\xi
\Gamma^\lambda{}_{\gamma\mu}\label{561}
\ea
where $\tilde \nabla$ denotes the covariant derivatives with respect to the
trasposed connection
$\tilde\Gamma^\lambda{}_{\gamma\mu}=\Gamma^\lambda{}_{\mu\gamma}$. Finally,
if we insert  \form{54}, \form{56} and \form{561} into \form{53} we end up
with:
\be
L'({K(\xi)})=\sqrt{g} \left\{G^{\mu\rho}  \pounds_\xi  g_{\mu\rho}+
2 g^{\sigma\gamma}
\T^\beta{}_{\lambda\sigma}
\pounds_\xi \Gamma^\lambda{}_{\gamma\beta}\right\}\, ds
\label{57}
\ee
where:
\be
\T^\beta{}_{\lambda\sigma}=T^\beta{}_{\lambda\sigma}-\delta^\beta_\sigma
T^\rho{}_{\lambda\rho}+\delta^\beta_\lambda T^\rho{}_{\sigma\rho}
\ee
is the
\emph{modified} torsion tensor; see \cite{Hehl}. Notice that the variational
Lagrangian of our initial theory turns out to be completely natural. Notice
also that
\form{57} is nothing but the variational Lagrangian ensuing form the
Einstein--Cartan Lagrangian density in the metric affine formalism:
\be
\L_{EC}=\sqrt{g} g^{\mu\nu} R_{\mu\nu}(j^1\Gamma)\label{EC}
\ee
where $\Gamma$ is the metric compatible connection \form{gammacomp}. Indeed
field equations ensuing from \form{EC} through a variational principle
\emph{\`a la} Palatini are nothing but  the coefficients in front of the Lie
derivatives in
\form{57} (except for a sign), i.e.
\ba
{\delta \L_{EC}\over \delta g_{\mu\rho}} &=&-\sqrt{g} \left\{R^{(\mu\rho)} -
{1\over 2}
g^{\mu\rho}
\, R
\right\}\label{EC1}\\
{\delta
\L_{EC}\over\delta\Gamma^\lambda{}_{\gamma\beta}}&=&-2\sqrt{g}g^{\sigma\gamma}
\T^\beta{}_{\lambda\sigma}
\ea 
\begin{Remark}{\rm We point out that  only the symmetric components of
the  Ricci tensor enter into  the Lagrangian \form{EC}  and in its associated
variational Lagrangian \form{57}. This fact justifies a posteriori the
condition \form{54}.\CVD}
\end{Remark}\vspace{1truecm}

 Notice, however, that in order to extract the
information  about the variation of conserved quantities  from the 
variational Lagrangian
\form{57} we have to take   into account that the linear connection
$\Gamma$ is a dynamical field
$\Gamma(j^1g, K)$ depending on the metric, its derivatives and the contorsion
tensor; see \form{gammacomp}. In other words (when implementing the first
step) the variational derivatives  of the Lagrangian 
\form{57} are to be performed with respect to the independent fields
$g_{\mu\nu}$ and
$K^\alpha{}_{\beta\mu}$ (or, equivalently, with respect to $g_{\mu\nu}$ and
$\T^\alpha{}_{\beta\mu}$).  After a straightforward calculation we get:
\be
\sqrt{g}\, G^{\mu\rho}  \pounds_\xi g_{\mu\rho}+
2 g^{\sigma\gamma}
\T^\beta{}_{\lambda\sigma}
 \pounds_\xi \Gamma^\lambda{}_{\gamma\beta}
=L_1+L_2+L_3\label{555}
\ee
where:
\ba
L_1&=&
\sqrt{g} G^{\mu\nu}(j^2g)  \pounds_\xi
g_{\mu\nu}\label{5551}\\
L_2&=& \pounds_\xi\left\{-\sqrt{g} g^{\mu\nu}(  K^\rho{}_{\sigma\rho}
K^\sigma{}_{\mu\nu} - K^\alpha{}_{\sigma\nu}
K_{\mu}{}^\sigma{}_{\alpha})\right\}\label{5552}\\
L_3&=&d_\beta\left\{-2\sqrt{g}\,\T^{\mu\beta\nu} \pounds_\xi g_{\mu\nu})
\right\}\label{5553}
\ea
The potential for $L_1$ has already  been computed in section
\ref{section4}  and is given  (apart from a factor $2k$) by formula \form{30}. 
 Moreover there is 
no contribution to the potential from $L_2$ since $L_2$ affects only the term
$\delta\tilde
\E$ in \form{current}. 
Indeed   we have ({\bf first step}):
\be
\delta L_2=d_\rho\left\{-\xi^\rho \delta\left(\sqrt{g} g^{\mu\nu}( 
K^\alpha{}_{\sigma\alpha} K^\sigma{}_{\mu\nu} - K^\alpha{}_{\sigma\nu}
K_{\mu}{}^\sigma{}_{\alpha})\right)\right\}
\ee
hence, see \form{deltaL'}:
\be
\F^\rho(L_2,\delta g, \delta K)=-\xi^\rho \delta\left(\sqrt{g} g^{\mu\nu}( 
K^\alpha{}_{\sigma\alpha} K^\sigma{}_{\mu\nu} - K^\alpha{}_{\sigma\nu}
K_{\mu}{}^\sigma{}_{\alpha})\right)
\ee
The latter expression is linear in $\xi$ so that, in the second step, no
further integration with respect to $\xi$ can be made, namely no divergence
term entering into the total potential comes out from it.

 The potential
 receives instead a contribution from the term $L_3$. To obtain it we have to
consider the part enclosed between the braces in \form{5553}, develop 
explicitly the Lie derivatives and then apply \form{FFTILDE2} with $h+r-1=1$.
The result turns out to be
\be
\U^{\beta\alpha}(L_3,\delta g, \delta \T)=\delta\left\{2\,\sqrt{g}(
\T^{[\alpha}{}_\rho{}^{\beta]}+ \T_\rho{}^{\alpha\beta})\xi^\rho
 \right\}\label{ult}
\ee
Hence the total potential for the variational Lagrangian \form{57} is given by
the sum of the purely metric potential \form{30} plus \form{ult}, namely:
\ba
\U^{[\beta \alpha]}&=&\delta \left\{ 2\sqrt{g}\Dg{}^{[\alpha}
\xi^{\beta]}\right\} + 2\sqrt{g}\, g^{\mu\nu}\,\delta
u^{[\alpha}_{\mu\nu}\,\xi^{\beta ]}\nonumber\\
&+&\delta\left\{2\,\sqrt{g}(
\T^{[\alpha}{}_\rho{}^{\beta]}+ \T_\rho{}^{\alpha\beta})\xi^\rho
 \right\}\label{DQEC}
\ea
where $u^\rho_{\mu\nu}=\gamma^\rho_{\mu\nu}-\delta^\rho_{(\mu}
\gamma^\sigma_{\nu)\sigma}$.\footnote{
The same result \form{DQEC} could be also achieved,  in a quite more
involved   way,  by inserting the expression 
 \form{Kosmann} of the Kosmann lift directly into 
\form{cumber} 
} We remark that, if the spin contribution
of the matter is zero, it holds true that 
$\T^\rho{}_{\lambda\sigma}\simeq0$. Hence,  expression \form{DQEC}
reproduces  on--shell exactly expression \form{30}. When the spin contribution
of matter is not vanishing it gives to  conserved
quantities a geometric contribution (i.e. relative to
the pure gravitational part of the theory)  which is produced by the torsion
terms in
\form{DQEC}. Notice however that the equation \form{39B} is an algebric
relation between torsion and spin matter meaning that, outside the
distribution of matter, there is no torsion, i.e. torsion does not propagate
(see \cite{Hehl}). The only way the spin of matter affects  conserved
quantities  outside the
distribution of matter (where \form{DQEC} collapses into \form{30})
is through  its influence on the metric tensor. 
\section{ Conclusions and perspectives}
We have described a  theory which  algorithmically defines the variation of
conserved quantities in the realm of gauge natural theories. The
formalism is developed through a two steps procedure  starting from the
variational Lagrangian  and  by making use of techniques related to the
Calculus of Variations in the jet bundle framework. We have tested  the
viability of the  results the theory predicts by applying them to
gravitational theories. In the classical Einstein's  metric formulation we
reproduced results already found elsewhere. Moreover, when dealing with ECKS
theory, viewed as a  gauge natural generalization of gravity, a preferred
lift of spacetime vector fields, namely the Kosmann lift,  is automatically
selected  by the formalism itself in order to restore naturality in the gauge
natural framework. We clearly welcome  this additional unlooked--for result
which has  turned up from the theory since it seems to endow   the Kosmann
lift, that in literature  was previously justified only on physical grounds
(see \cite{Ultimo,BTZ,Matteucci}), also with  a good mathematical  reason of
existence.

Moreover we point out that the left hand side of equations \form{39A} and
\form{39B} can be easily generalized to encompass any spacetime
dimension $D$, $D>2$. In particular, their generalization  to odd dimensions 
reproduces the field equations content of Chern-Simon
theories of the group $SO(2, D-1)$ or
$SO(1,D)$, see \cite{CF}. The case $D=3$ has been analyzed in
\cite{Ultimo} following  the spirit of the formalism  presented here. We
believe that it will be  worth investigating also the case $D=5$ (as well as 
any larger  dimension), mainly because of its relationship with the
Gauss--Bonnet generalization of Einstein's gravity, a generalization which
seems to be relevant in effective theories ensuing from the  low--energy limit
of string theory (see \cite{stringa}).

We also stress that our recipe is well suited to be generalized to a more
general class of symmetries other than those induced just by projectable vector
fieds on the structure bundle. 
Indeed the  fundamental requirement we made  on the Lie derivatives of the
fields is that they can be expanded as a linear combination of the
independent parameters
 $\xi^\mu$ and
$\xi^{A}$ together with their derivatives up to a  fixed (finite)
order. This is in fact the assumption which makes the
formalism algorithmically well--defined (see expression \form{RRR} and
\form{fff}). No restriction  has been instead imposed on  the coefficients of
the aforementioned expansion which, in principle, can depend not only on the
field together with  their first order derivatives, but also on an arbitrary
(but finite) number of  derivatives. Namely,  symmetries   with respect to
generalized vector fields (see
\cite{general,Olver} are not out of the scope of the formalism here developed.
Thereby we guess that the formalism is  in good position to deal with   the
so--called generalized symmetries, e.g. supersymmetries and BRST
transformations; see
\cite{Opava,super}.

\section{Acknowledgments}
  We are grateful to G.\ Allemandi,
L.\ Fatibene  and M.\ Godina of the
University of Torino for useful 
discussions and suggestions on the
subject.


\end{document}